# Real space imaging of one-dimensional standing waves: Direct evidence for a Luttinger liquid

Jhinhwan Lee[1], S. Eggert[2], H. Kim[1], S.-J. Kahng[3], H. Shinohara[4] & Y. Kuk[1,*]

[1]Department of Physics and Center for Science in Nanometer Scale,
Seoul National University, Seoul, 151-742, Korea

[2]Department of Physics, University of Kaiserslautern, Kaiserslautern, D67663 Germany

[3]Department of Physics, Korea University, Seoul 136-701, Korea

[4]Department of Chemistry, Nagoya University, Nagoya 464-8602, Japan

*Corresponding author. Email: ykuk@phya.snu.ac.kr

**Electronic standing waves with two different wavelengths were directly mapped near one end of a single-wall carbon nanotube (SWNT) as a function of the tip position and the sample bias voltage with high-resolution position-resolved scanning tunneling spectroscopy (PR-STS). The observed two standing waves caused by separate spin and charge bosonic excitations are found to constitute direct evidence for a Luttinger liquid (LL). The increased group velocity of the charge excitation, the power-law decay of their amplitudes away from the scattering boundary, and the suppression of the density of states (DOS) near the Fermi level were also directly observed or calculated from the two different standing waves.**

In two- and three-dimensional metals, electrons near the Fermi level are scattered around defects or impurities, forming electronic standing waves [1,2] – this is known as Friedel oscillation. The interacting electrons are well described by the Fermi liquid theory of non-interacting quasi-particles. In one-dimensional (1D) metals, standing waves of the screening electrons near the Fermi level can also be produced around defects. However, even weak electron-electron interaction can destroy the picture of non-interacting quasi-particles. These correlated 1D electrons are better described by the Luttinger liquid (LL) theory [3]. The main consequences of the LL theory are: 1) the spin and charge degrees of freedom of electrons are separated, 2) the charge velocity ($v_c$) is larger than the Fermi velocity ($v_F$) while the spin velocity ($v_s$) is close to $v_F$, and 3) the DOS is expected to exhibit a power-law suppression $N(\omega) \sim |\omega|^{\alpha}$ near the Fermi level ($|\omega| << E_F$) [4–6]. In the presence of a potential wall, a recent LL theory of scanning tunneling spectroscopy predicted the possibility of directly observing: 1) 1D standing waves in the position dependent DOS, 2) energy-dependent, multiple standing waves caused by increased group velocity of the charge excitation ($v_c$), and 3) a power-law decay of the standing waves away from the barrier [7]. It was indicated that real-space imaging of the multiple standing waves in a long 1D wire, such as metallic single-wall carbon nanotube (SWCNT), can only be described by separate bosonic excitations.

There have been several position-averaged experimental observations of LL behaviours in 1D electron systems: 1) The edge tunneling into a two-dimensional electron gas (2DEG) in the fractional quantum hall regime showed power-law behaviours of the current-voltage and the conductance-temperature characteristics [8,9]; 2) The momentum-resolved electron tunneling between two nanowires made by the cleaved-edge overgrowth method revealed the dispersion curves in each nanowire and some signatures of DOS suppression near Fermi-level and charge-spin separation [10-12]; 3) The photoemission spectra of carbon nanotube bundles [13] showed the power-law behaviour of the position- and angle-averaged electronic density of states; 4) The conductance of a metallic carbon nanotube showed power-law dependence on the temperature and on the bias voltage [14]. These experimental results, however, did not present the real space images of an LL but demonstrated fitting to the functional dependences of the experimental variables on the theoretical model.

In this report, we directly imaged the standing waves with two different wavelengths formed near one end of a SWCNT on Au(111) surface with atomic resolution and present four other items of direct evidences for an LL. We chose a metallic SWCNT because it is an ideal 1D wire of an LL with a linear dispersion near the Fermi level [15, 16]. The experimental details are described elsewhere [17]. Briefly, we deposited SWCNTs on an atomically clean Au(111) surface and then performed scanning tunneling microscope (STM) topography and PR-STS at 4.7K in ultra-high vacuum ($<1\times10^{-10}$ Torr). We could routinely observe clean SWCNTs on atomically clean Au(111) surfaces with the herring-bone structure.

Figure 1a is an STM topography of an SWCNT whose index is identified as (19, 7) and whose length is >1.1 μm. Figure 1b is the corresponding $dI/dV(r,V)$ map as functions of energy ($V$) and distance ($r$) from one end of the nanotube along the middle of the SWCNT. This $dI/dV(r,V)$ map shows the positional variation of the electron density of states. Two modulations are visible in Fig. 1b: 1) energy-resolved Friedel oscillations (fine ripples) with spatial wavelengths (0.3 ~ 0.4 nm) comparable to the lattice constant (0.25 nm) of the SWNT, 2) additional modulations with wavelength >~ 2 nm in the positive bias region. At the end of the SWNT, a potential wall is created by the SWCNT-vacuum barrier, illustrated as the blue area in Figs. 1b and 1c. The standing waves in the $dI/dV(r,V)$ image (energy-resolved Friedel oscillations of electron density of states) are caused by the interference of the incoming and scattered waves. The 1D energy dispersion of the charge waves gives the bias voltage dependence of the wavelengths as illustrated on the right side of Fig. 1b and 1c.

Figure 1c is a simulated image of the $dI/dV(r,V)$ map based on the LL theory [7]. We have employed a Luttinger parameter of $g$ = 0.55 (i.e., $v_F$ = 0.55 $v_c$) and $v_s \approx v_F$. We have also used multiple dispersion branches with Fermi-points at $k_F$ = ±11.9 nm$^{-1}$ and ±8.8 nm$^{-1}$, which correspond to a (19,7) nanotube in an extended zone scheme. The first 8.8 nm$^{-1}$ branch, the reflected mode at negative energies of the second 8.8 nm$^{-1}$ branch and the reflected mode of the second 11.9 nm$^{-1}$ branch of a (19,7) nanotube are not included in the simulations since they appear to be suppressed in the experimental data. This may be due to some symmetry in the structure at the end at the nanotube, which determines the scattering [18, 19]. In Fig. 1c, two modulations, fine ripples and additional modulation, are well reproduced with the above-given parameters. As suggested earlier, our direct imaging of multiple standing waves, caused by the increased group velocity of charge excitation ($v_c$), is the unambiguous proof of validity of LL theory.

The theory also predicted that the peak energy levels of the additional modulations scale with the distance from the scattering boundary. Figure 2 shows the energy-dependent spectroscopy ($dI/dV(V)$) data at different tip positions near one end of the SWCNT. The overall features (including the positions of the peaks) of the experimental data (solid curves) show good agreement with the simulation results of the LL theory (dashed curves) with $g$ = 0.55. The energy levels of the peaks and their differences decrease with increasing distance ($r$) from the end, scaling according to the relation $r \times V$ = const. in both the experimental and the theoretical data (the first piece of supporting evidence). The electron density of states (*dI/dV*) also decreases as $V \to 0$, showing the suppression of the local density of states near the Fermi level. That is the second piece of supporting evidence for an LL. We believe that the multiple peaks of super-modulations in Fig. 2 are not related to the Coulomb blockade effect, because the energy level spacing of Coulomb blockade peaks is directly related to the charging energy (hence the geometry of the quantum dot itself) and should not vary continuously as a function of the tip position or the coupling strength [6].

By taking a Fourier transformation of each horizontal line in Figs. 1b and 1c, we can construct *dI/dV*(*k,V*) maps, similar to the dispersion relation between *E*(*V*) versus *k*. Since the imaged standing waves represent the charge density waves, their *k* components are twice the *k* components of the electronic wavefunctions allowed by the 1D energy dispersion. Figures 3a and 3b show the corresponding experimental and theoretical *dI/dV*(2*k,V*) maps superposed with the 1D energy dispersion curves (red dashed curves in Fig. 3a and 3b) of a (19,7) nanotube calculated using a tight-binding (TB) method. The bright colored spots near the Fermi level in Fig. 3a correspond to the corners of the 1st and 2nd Brillouin zones projected on the tube axis. The sloped curves near $k = \pm 8.8$ nm$^{-1}$ and $k = \pm 11.9$ nm$^{-1}$ indicated by the yellow arrows make the largest contribution to the ripples in Fig. 1b. From their slopes we can calculate the group velocity of the electron waves. Although the sloped curves look similar to the TB-calculated energy dispersion curves, their slopes are significantly steeper, corresponding to a larger charge-mode group velocity of $v_c \approx v_s/0.55$, where $v_s \approx v_F \approx 8.1 \times 10^5$ m/s. This is the third piece of supporting evidence for an LL in a SWCNT, confirmed in this study. The increased charge velocity is evidently visible in the theoretical simulation in Fig. 3b, as observed in the experimental data in Fig. 3a. The theoretical simulation shows that the spin-related modulation branch is considerably weaker, which may explain why we could not clearly resolve them in the experimental data.

The LL theory predicts that the decay of the standing wave amplitude follows a power law of $A \sim r^{-\alpha}$ at the far-field region ($r \gg v/V$) and at high bias voltage ($|V| \gg \Lambda \sim 0.1$ V) where $\alpha = (2+g+g^{-1})/8$ [7]. In Fig. 4, we plot the amplitude of the standing waves as a function of $r$ in the region of 5.6 nm ~ 15.7 nm, averaged over the bias voltage of $-1.5$ V ~ $-1.0$ V. The data points at far-field region ($r > \sim 9$ nm) are well fitted by the theoretical slope (solid line) of $\alpha \sim 0.55$ for $g \sim 0.55$ (the fourth piece of supporting evidence for an LL). The slope (0.79) of the globally fitted (dotted) line is somewhat greater than the theoretical value due to some anomalous behaviour in the near-field region.

The observed value of $g \sim 0.55$ is much larger than the value of $g \sim 0.3$ for a weakly screened nanotube predicted by theory [6] and for nanotubes on the oxide layer in a transport experiment [14]. This can be explained by the screened Coulomb interaction of the 1D nanotube electrons due to the underlying metallic substrate. If the reduced screening length due to the substrate is $R_s$ and the radius of the nanotube is $R$, the charge velocity is given by $v_c = [v_F\{v_F+(8e^2/\pi\hbar)\ln(R_s/R)\}]^{1/2}$ [6]. With $8e^2/\pi\hbar \approx 6.88 v_F$, we have an estimated ratio of $R_s/R \approx 1.4$ for g = $v_F/v_c$ = 0.55, or $R_s \approx 1.1$ nm assuming $R$ = 0.75 nm. This shows that the underlying Au(111) substrate significantly reduced the effective Coulomb interaction range of the 1D electrons in the nanotube and increased the Luttinger parameter accordingly.

In summary, we directly imaged electron standing waves with two wavelengths at one end of a nanotube on a metallic substrate with atomic resolution using the method of position-resolved scanning tunneling spectroscopy. The real-space and Fourier-transformed spectroscopic data and the power-law decay of the standing wave intensity agree well with the LL theory. The screening effect due to the metallic substrate seems to increase the Luttinger $g$ parameter significantly. The method presented here, if applied to various 1D nano-conductors on different substrates and with varying contact conditions, may enhance our understanding of correlated 1D electron systems and their applicability to future electronic devices.

This work was supported by the Korean Ministry of Science and Technology through Creative Research Initiatives Program and the Future Program on New Carbon Nano-Materials by the Japan Society for the Promotion of Science.

**References and Notes**

[1] M.F. Crommie, C.P. Lutz, D.M. Eigler, *Science* **262**, 218-220 (1993).

[2] Y. Hasegawa, Ph. Avouris, *Phys. Rev. Lett.* **71**, 1071 (1993).

[3] J.M. Luttinger, *Phys. Rev.* **119**, 1153 (1960).

[4] A.O. Gogolin, A.A. Nersesyan, A.M. Tsvelik, *Bosonization and strongly correlated systems* (Cambridge Univ. Press, New York City, 1998).

[5] J. Voit, *J. Phys. Condens. Matter* **5**, 8305 (1993).

[6] C. Kane, L. Balents, M.P.A. Fisher, *Phys. Rev. Lett.* **79**, 5086 (1997).

[7] S. Eggert, *Phys. Rev. Lett.* **84**, 4413 (2000).

[8] X.G. Wen, *Phys. Rev. Lett.* **64**, 2206 (1990).

[9] A.M. Chang, L.N. Pfeiffer, K.W. West, *Phys. Rev. Lett.* **77**, 2538 (1996).

[10] A. Yacoby *et al.*, *Phys. Rev. Lett.* **77**, 4612 (1996)

[11] O.M. Auslaender *et al.*, *Science* **295**, 825 (2002).

[12] Y. Tserkovnyak *et al.*, *Phys. Rev. Lett.* **89**, 136805-1 (2002).

[13] H. Ishii *et al.*, *Nature* **426**, 540 (2003).

[14] M. Bockrath *et al.*, *Nature* **397**, 598 (1999).

[15] S.G. Lemay *et al.*, *Nature* **412**, 617 (2001).

[16] M. Ouyang, J.L. Huang, C.M. Lieber, *Phys. Rev. Lett.* **88**, 066804-1 (2002).

[17] J. Lee *et al.*, *Nature* **415**, 1005 (2002).

[18] For example, Choi et al. (H. J. Choi *et al.*, *Phys. Rev. Lett.* 84, 2917 (2000)) shows that the electron scattering from a pentagon-heptagon-pair defect can have strong directionality (Fig. 3, bottom right image), leaving nearly zero amplitude in the tube axis direction. If such directionality could be valid for a mode (a branch of dispersion curve) in the entire voltage range of STS measurement, the mode will be totally invisible in FT-STS measurement.

[19] Even with the tip condition that gives the highest resolution spectroscopy, the energy-resolved Friedel oscillations could be resolved only for r > ~2 nm. A simple Luttinger liquid model with the assumption of infinite potential barrier may not be appropriate very close to the end of the tube due to the strong singularity at the scattering boundary.

**Figure Captions**

**FIG. 1:** The standing waves of electrons scattered at one end of a nanotube peapod. An STM topographic image (a) acquired at sample bias voltage of 1 V and the corresponding differential conductance (density of states) map (b) as a function of the tip position (abscissa) and the sample bias voltage (ordinate). The horizontal scan range of (a) and (b) is 8.3 nm. (c) A simulated plot of the $dI/dV(r,V)$ map based on the LL theory using $g = 0.55$, $v_s = 0.55v_c \approx v_F$, and Fermi points at $k_F = \pm 11.9$ nm$^{-1}$ and $\pm 8.8$ nm$^{-1}$.

**FIG. 2:** Measured (solid curves) and simulated (dashed curves) electron density of states ($dI/dV$) near one end of the nanotube at tip positions indicated on the right. Measured (filled) and simulated (unfilled) peak positions are marked by triangles. Black triangles indicate the suppression of the density of states near the Fermi level. Each pair of curves is shifted by one unit for better visibility.

**FIG. 3:** The measured (a) and the simulated (b) Fourier-transformed maps of $dI/dV$ (density of states) plotted as a function of $k$, the electron momentum, and $V$, the sample bias voltage, respectively. The ovals near the Fermi level indicate the corners of the Brillouin zones nearest (green) and second nearest (orange) to the $\Gamma$ point ($k = 0$). The red dotted curves are the 1D energy dispersion of a (19,7) nanotube calculated using a tight-binding method.

**FIG. 4:** A log-log plot of the standing wave amplitude versus the distance from one end of the nanotube in the distance range of 5.6 nm $< r <$ 15.7 nm. Each data point is averaged over the bias range of -1.5 V $< V <$ -1.0 V to reduce the contribution of the atomic lattice corrugation. The solid line shows the theoretically expected slope (~ 0.55) that fits fairly well with the data points at far-field region ($r >\sim 9$

nm). The dashed line shows the global (both near- and far-field) power-law fit with the slope of ~ 0.79.

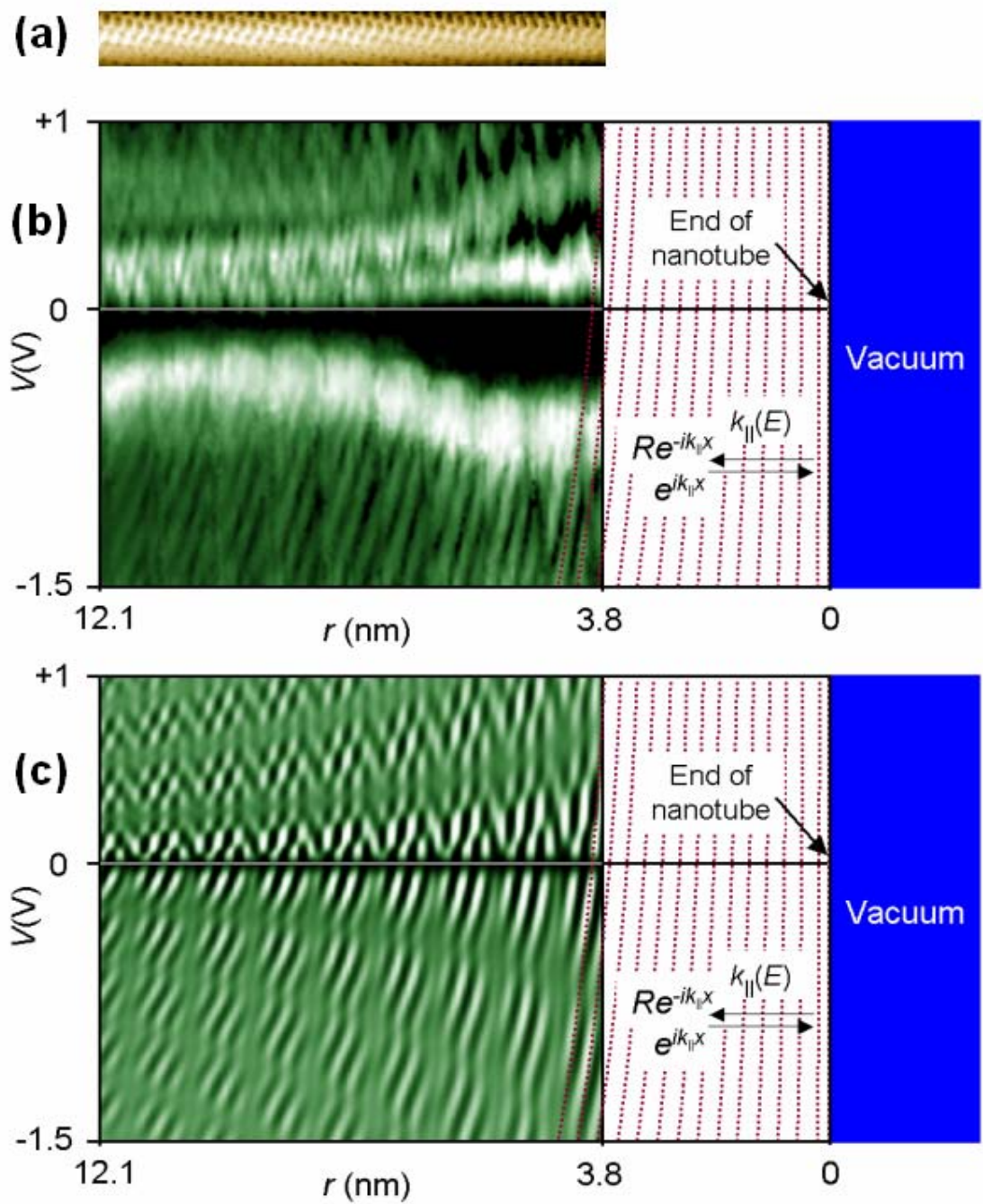

Fig. 1 Lee

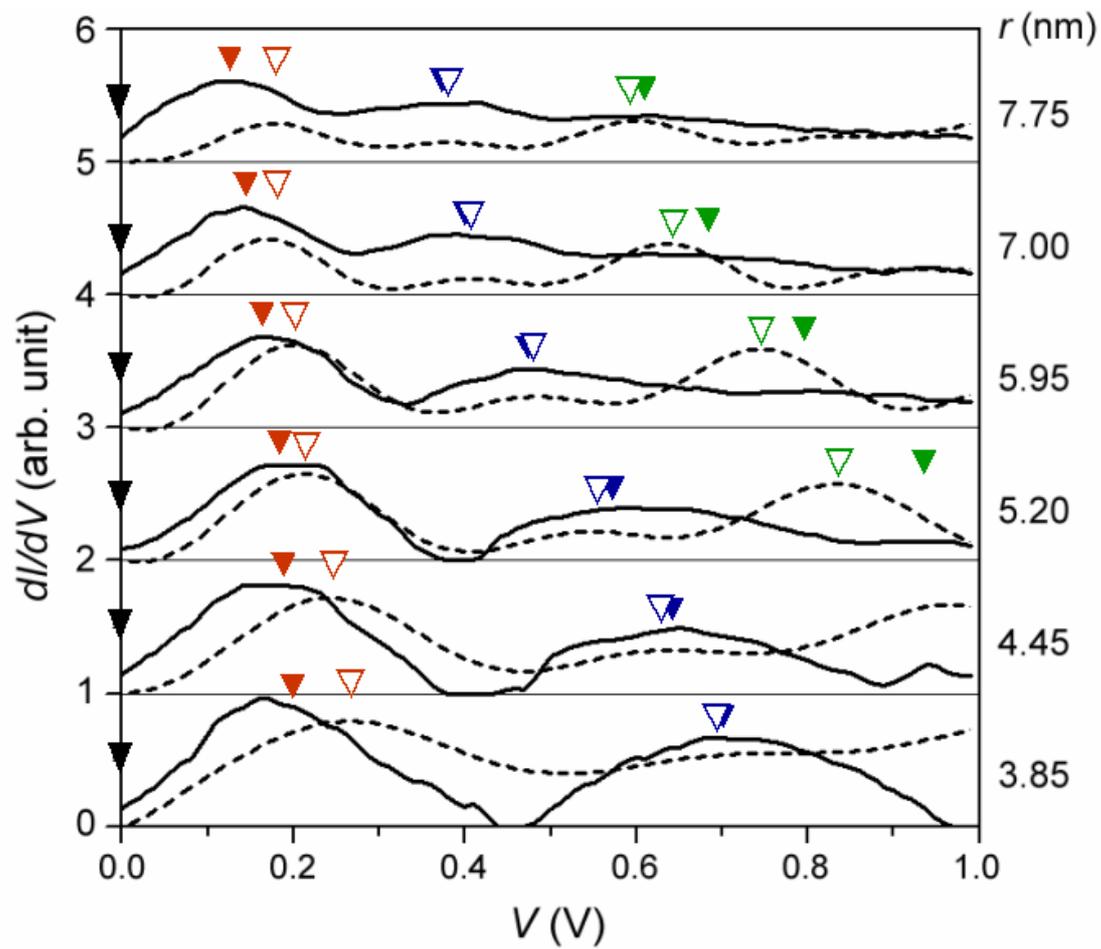

Fig. 2 Lee

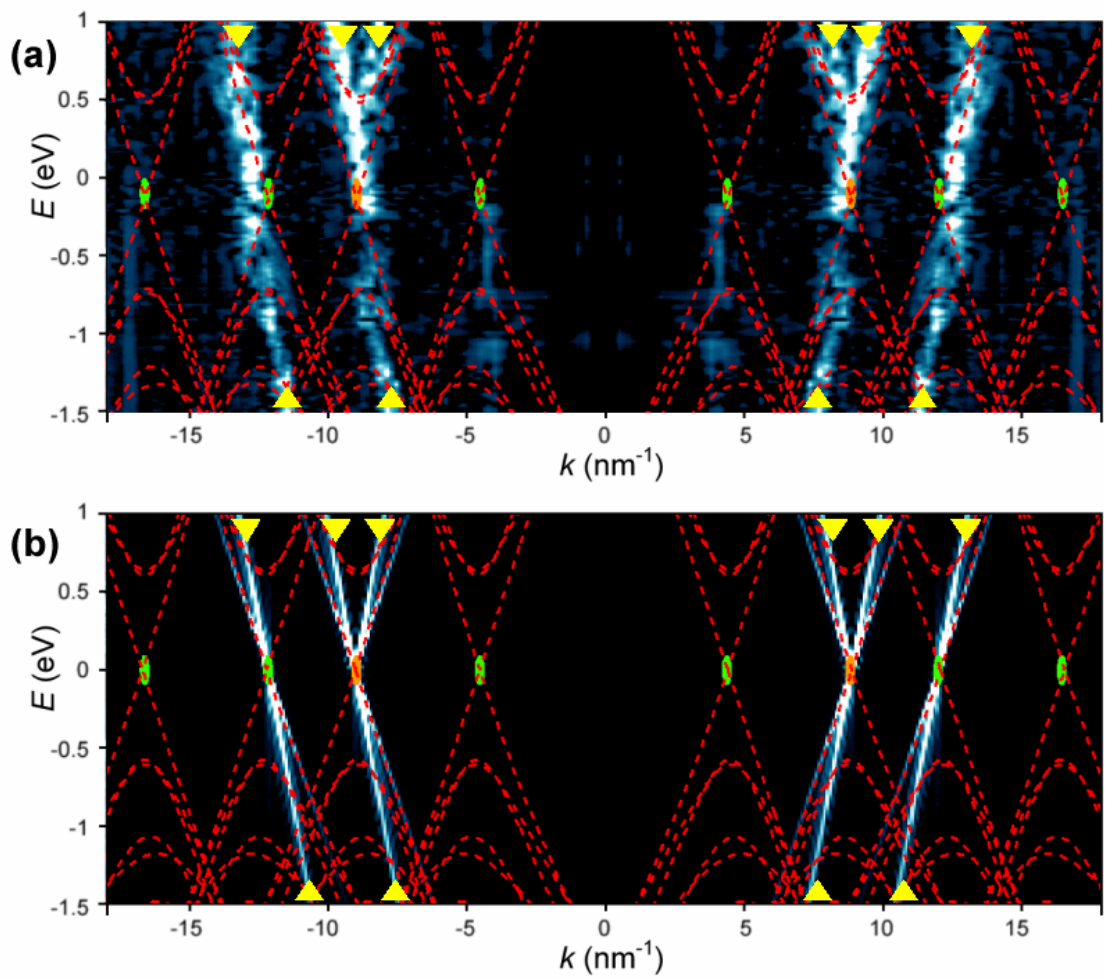

Fig. 3 Lee

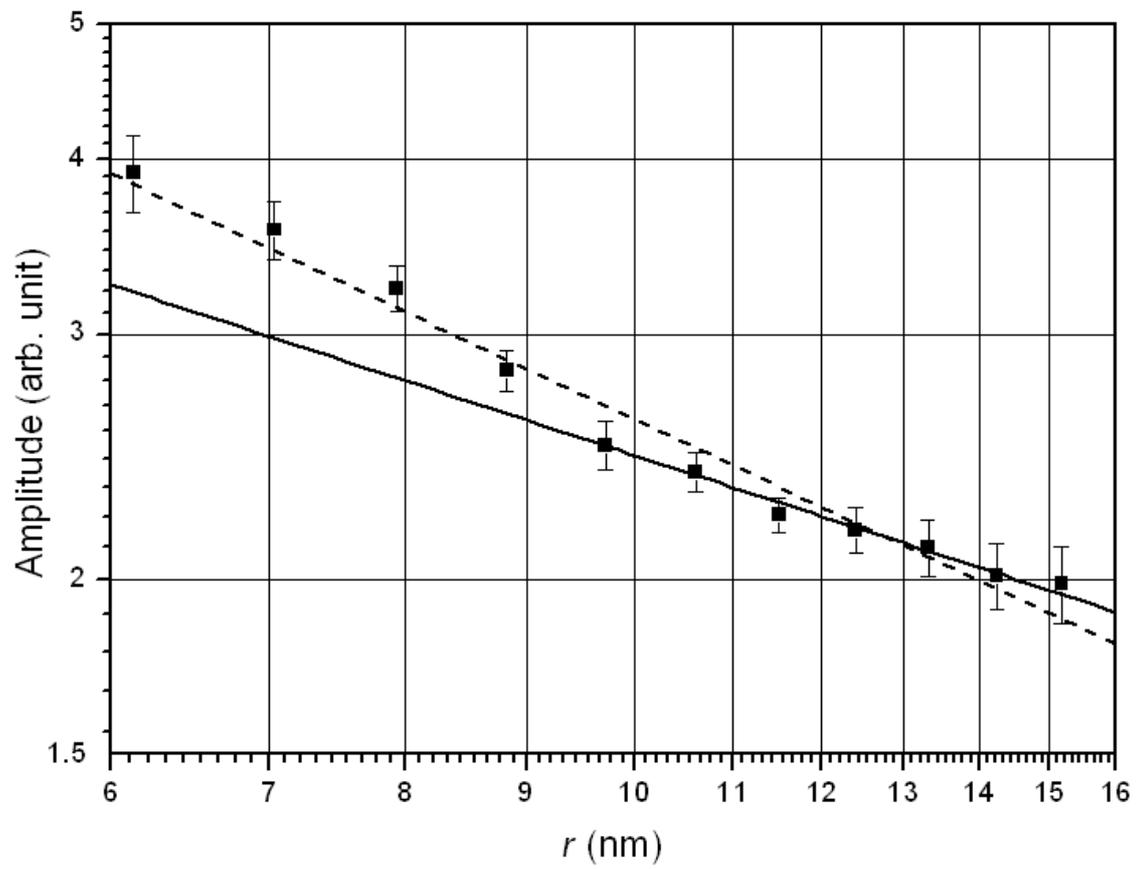

Fig. 4 Lee